
\magnification=\magstep1
\hsize=16.0  true cm
\baselineskip=12pt
\parskip=0.2cm
\parindent=1cm
\raggedbottom

\def\pp{\parshape 2 0truecm 16truecm 1truecm 15truecm}
%
\def\apjref#1;#2;#3;#4 {\par\pp#1,  #2,  #3, #4 \par}
%
%
\def\oldapjref#1;#2;#3;#4 {\par\pp#1, {\it #2}, {\bf #3}, #4. \par}
%
%

\def\section#1{\bigskip\goodbreak\centerline{\bf #1}}
%
%
\def\subsection#1{\goodbreak\noindent{\bf #1}}
\def\subsubsection#1{{\noindent \it #1}}
\def\ltsima{$\; \buildrel < \over \sim \;$}
\def\simlt{\lower.5ex\hbox{\ltsima}}
\def\gtsima{$\; \buildrel > \over \sim \;$}
\def\simgt{\lower.5ex\hbox{\gtsima}}
%
%



\newcount\notenumber
\notenumber=1
\def\note#1{\footnote{$^{\the\notenumber}$}{#1}\global\advance\notenumber by 1}

\newcount\eqnumber
\eqnumber=1
\def\new{{\rm(\chaphead\the\eqnumber}\global\advance\eqnumber by 1}
\def\ref#1{\advance\eqnumber by -#1 (\chaphead\the\eqnumber
     \advance\eqnumber by #1 }
\def\last{\advance\eqnumber by -1 {\rm(\chaphead\the\eqnumber}\advance
     \eqnumber by 1}
\def\eq#1{\advance\eqnumber by -#1 equation (\chaphead\the\eqnumber
     \advance\eqnumber by #1}
\def\eqnam#1#2{\immediate\write1{\xdef\ #2{(\chaphead\the\eqnumber}}
    \xdef#1{(\chaphead\the\eqnumber}}

\newcount\fignumber
\fignumber=1
\def\nfig{\the\fignumber\ \global\advance\fignumber by 1}
\def\nfiga#1{\the\fignumber{#1}\global\advance\fignumber by 1}
\def\rfig#1{\advance\fignumber by -#1 \the\fignumber \advance\fignumber by #1}
\def\fignam#1#2{\immediate\write1{\xdef\
#2{\the\fignumber}}\xdef#1{\the\fignumber}}

\newbox\abstr
\def\abstract#1{\setbox\abstr=\vbox{\hsize 5.0truein{\par\noindent#1}}
    \centerline{ABSTRACT} \vskip12pt \hbox to \hsize{\hfill\box\abstr\hfill}}




\def\s{\ifmmode \widetilde \else \~\fi}
\def\={\overline}

\def\spose#1{\hbox to 0pt{#1\hss}}

\def\lta{\mathrel{\spose{\lower 3pt\hbox{$\mathchar"218$}}
     \raise 2.0pt\hbox{$\mathchar"13C$}}}
\def\gta{\mathrel{\spose{\lower 3pt\hbox{$\mathchar"218$}}
     \raise 2.0pt\hbox{$\mathchar"13E$}}}
\def\Dt{\spose{\raise 1.5ex\hbox{\hskip3pt$\mathchar"201$}}}	
\def\dt{\spose{\raise 1.0ex\hbox{\hskip2pt$\mathchar"201$}}}	


\def\=={\equiv}

\def\dotsfill{\leaders\hbox to 1em{\hss.\hss}\hfill}









\rm		
\def\Atoday{\ifcase\month\or
  January\or February\or March\or April\or May\or June\or
  July\or August\or September\or October\or November\or December\fi
  \space\number\day, \number\year}
\def\Etoday{\number\day\space\ifcase\month\or
  January\or February\or March\or April\or May\or June\or
  July\or August\or September\or October\or November\or December\fi
  \space\number\year}

\centerline {\bf GOOD ABUNDANCES FROM BAD SPECTRA: II. }
\centerline {\bf APPLICATION AND A NEW STELLAR COLOR-TEMPERATURE CALIBRATION}
\medskip
%
%
 \bigskip
 \centerline { J.~BRYN JONES}
 \centerline {Institute of Astronomy, Madingley Road, Cambridge CB3 0HA,
England, UK}
\smallskip
\centerline {and}
\smallskip
\centerline {Department of Physics and Astronomy,\footnote{$^1$}{Present
 Address}
 University of Wales,}
\centerline{College of Cardiff, PO Box 913, Cardiff, CF4 3TH,
Wales, UK}
\smallskip
\centerline {Electronic mail: jbj@astro.cf.ac.uk}
 \bigskip
 \centerline {ROSEMARY F.G.~WYSE}
 \centerline {Center for  Particle Astrophysics, University of California,
Berkeley, CA 94720, USA}
\smallskip
 \centerline {and}
\smallskip
 \centerline {Institute of Astronomy, Madingley Road, Cambridge CB3 0HA,
England, UK}
\smallskip
\centerline {and}
\smallskip
\centerline {Department of Physics and Astronomy,\footnote{$^2$}{Permanent
 Address}
The Johns Hopkins University,}
\centerline {Baltimore, MD 21218, USA}
\smallskip
\centerline {Electronic mail: wyse@tardis.pha.jhu.edu}
 \bigskip
 \centerline {GERARD GILMORE}
 \centerline {Institute of Astronomy, Madingley Road, Cambridge CB3 0HA,
England, UK}
\smallskip
\centerline {Electronic mail: gil@ast.cam.ac.uk}
 \bigskip
Running Head : Good Abundances from Bad Spectra

 \vfill\eject
 \centerline {\bf ABSTRACT}

\medskip

Stellar spectra derived from current multiple-object fiber-fed
spectroscopic radial-velocity surveys, of the type feasible with,
among other examples, AUTOFIB, 2dF, HYDRA, NESSIE, and the Sloan
survey, differ significantly from those traditionally used for
determination of stellar abundances. The spectra tend to be of
moderate resolution (around $1\,${\AA}) and signal-to-noise ratio
(around 10-20 per resolution element), and cannot usually have
reliable continuum shapes determined over wavelength ranges in excess
of a few tens of Angstroms. Nonetheless, with care and a calibration
of stellar effective temperature from photometry, independent of the
spectroscopy, reliable iron abundances can be derived.

We have developed techniques to extract true iron abundances and
surface gravities from low signal-to-noise ratio, intermediate
resolution spectra of G-type stars in the 4000--$5000\,${\AA}
wavelength region.  The theoretical basis and calibration using
synthetic spectra are described in detail in another paper (Jones,
Gilmore and Wyse 1995). The practical application of these techniques
to observational data, which requires some modification from the ideal
case of synthetic data, is given in the present paper.  An
externally-derived estimate of stellar effective temperature is
required in order to constrain parameter space sufficiently; a new
derivation of the $V-I$ -- effective temperature relation is thus an
integral part of the analysis presented here. We have derived this
relationship from analysis of available relevant data for metal-poor G
dwarfs, the first such calibration.  We test and calibrate our
techniques by analysis of spectra of the twilight sky, of member stars
of the cluster M67, and of a set of field stars of known metallicity.
We show that this method, combined with our new color-temperature
calibration, can provide true iron abundances, with an uncertainty of
less than $ 0.2$ dex over the range of metallicity found in the
Galactic thick and thin disks, from spectra obtained with fiber-fed
spectrographs.

\vfill\eject

\section{1. INTRODUCTION }

Multi-object optical fiber-coupled spectrographs are increasingly
being used for radial-velocity surveys of faint objects. Such surveys
generate very many spectra, each of which is without reliable
continuum flux calibration (due to the difficulty of atmospheric
dispersion correction with fibers), and of lower resolution and
signal-to-noise ratio than are typically used for stellar abundance
determinations.  Nonetheless, useful chemical abundance data can be
derived from such spectra.  We have described in Paper I (Jones,
Gilmore and Wyse 1995) a method developed to measure iron abundances
and surface gravities from relatively low signal to noise ratio,
intermediate-resolution (around $1\,${\AA}) spectra of G-type stars, in
the 4000--$5000\,${\AA} wavelength region.  The details of the
theoretical framework and its calibration by means of the analysis of
synthetic spectra are given in Paper I.  The present paper concerns
the practical application to observational data. Gilmore, Wyse and
Jones (1995) apply these methods to a sample of faint ($15^{^m}<{
V}<18^{^m}$) F- and G-type stars a few kiloparsecs below the Galactic
plane to determine the thick disc chemical abundance distribution.

The strengths of absorption lines in the atmosphere of a star depend
on the effective temperature, the stellar surface gravity, and the
abundances of different chemical elements.  In principle, all these
parameters may be measured, given high enough quality data.  Lower
signal-to-noise ratio data require some compromises. For example, it is
potentially dangerous to attempt to solve for both temperature and
metallicity for late-type stars, since errors in these two parameters
are highly correlated, due to the strong sensitivity of metallic line
absorption to both parameters.  The effective temperature of a star
may be estimated from non-spectroscopic data, such as $V-I$ colors,
leaving gravity and elemental abundances as the determining parameters
of the strengths of absorption features.  Defining absorption indices
of different sensitivities to abundance and gravity allows a solution
for the two parameters to be obtained.  This is the philosophy behind
the approach advocated here.  The problem may be further simplified if
one can estimate the surface gravity of the star, for example from an
expectation of the evolutionary state, reducing the analysis to a
study of the relationship between line strength and elemental
abundance.  This allows poorer quality data to provide chemical
abundance estimates.

Spectroscopic indices sensitive to iron abundance and gravity have
been defined from a set of narrow (few -- several~{\AA} wide)
wavelength bands, with adjacent comparison bands.  The theoretical
calibration of this technique, presented in paper I, indicated
that, given a photometrically determined effective temperature, it
should be possible to derive relations between iron abundance and
gravity from each individual absorption line index, and that each of
these two stellar parameters may be determined, provided that the
noise in the intensity data is not too great. Further, a single
abundance indicator may be defined for the reduction of noisier data;
with the input of values for both the surface gravity and the
effective temperature, this single indicator is able to provide useful
iron abundance information from spectra having signal-to-noise ratios
as low as 10 per resolution element.

\section{2. DETERMINATION  OF IRON ABUNDANCE}

\subsection{2.1 The Basic Analysis Methods}

As discussed in paper I, 80 flux bands were chosen as the bases of
metallicity- and gravity-sensitive absorption-line indices, akin to
equivalent widths.  These are given in Table 1 (metallicity) and Table
2 (gravity), together with the set of eleven metallicity and
metallicity-gravity indices which have been defined from these (Table
1), and five ionic gravity indices (Table 2).  The technique for
estimating the iron abundance of a given star rests upon the
comparison of the values of these indices, or a suitable combination
of them, measured from an observed spectrum, with those of synthetic
spectra with effective temperature (and later, gravity) chosen to
bracket the star under study.

Synthetic spectra were computed assuming LTE for a range in stellar
parameters -- metallicity, gravity and effective temperature --
covering the grid given in Table 3, giving a total of 100 stellar
models.  These are scaled from the standard Holweger-M\"{u}ller (1974)
solar model atmosphere, adopting solar ratios for the relative
abundances of the heavy elements; any effects of non-solar
element ratios are taken account of by the empirical calibration below.
The data covered the wavelength ranges 4025--$4090\,${\AA}$\,,$
4500--$4690\,${\AA} and 4870--$5000\,${\AA}$\,,$ these sections being
selected as the most promising for iron abundance indices in the blue
region of the spectrum, whilst also being free of substantial molecular
line absorption.  All atomic lines listed in the Moore, Minnaert and
Houtgast (1966) solar line compilation were included in the synthesis
calculations.  Oscillator strengths were computed from the solar
spectrum for most lines, supplemented by selected accurate laboratory
data. Details of the synthesis process, including the choice of solar,
stellar and atomic data, are given in Paper I and in Jones (1991).

The fluxes in the narrow wavelength bands are somewhat sensitive to
the resolution of the spectrum, as a result of the extent to which
features at the edges of the bands are included.  In order to allow
the analysis methods to be used for the reduction of spectra of
differing resolutions, flux data were computed appropriate to a range
of resolution, from $1.0\,${\AA} to $2.5\,${\AA} full-width at
half-maximum in $0.1\,${\AA} intervals.  Each theoretical spectrum was
convolved with a Gaussian profile of an appropriate width to represent
instrumental effects adequately. The validity of this broadening
function, and of its width, were gauged by the study of calibration
arc-lamp lines. A fixed resolution was adopted across the spectrum;
the actual variation of the resolution across the observed spectra of
interest was generally a few percent (see section 4.1 below), so this
assumption is reasonable.  Residual fluxes, where `residual' means
the ratio of the observed value to that which would occur in the
absence of absorption, were calculated by integrating the residual
intensities over wavelength across each flux band.  Each set of 8000
synthetic fluxes corresponding to a particular resolution was stored
in a separate data file; the analysis methods can be tailored to the
resolution of a particular observed spectrum by the choice of the
synthetic flux data file.

A difficulty with multi-object fiber-fed spectrographs, which
complicates comparison of synthetic and observed fluxes, is the
presence of scattered light in the spectrograph and detector, adding a
locally-variable background signal to the reduced spectra. Care is
essential to correct reliably for this scattered light. For the
spectra of relevance here the optimum sky background subtraction
algorithm applies a local scattering correction (see Wyse and Gilmore
1992), obviating the need for any special processes to handle
scattered light in the abundance analysis.

\subsection {2.2  Evaluation of the  Indices}

An observed spectrum is calibrated in wavelength, the stellar radial
velocity is determined by cross-correlation against templates of known
velocity (the present data provide velocities accurate to 5--10 km/s),
the spectrum is corrected to a laboratory rest frame, then normalised
to a constant continuum level.  The continuum variation is removed by
fitting distinct low-order polynomials to sections of the spectra,
followed by a smoothing of the final
fit. The errors of this procedure are usually less than 10~\% in
intensity over 100~{\AA}\ at short wavelengths, where molecular bands
are strong, and considerably less than this at longer wavelengths.
The errors in the normalisation will tend to be greatest in spectra of
stars which experience strong absorption, in effect in cool stars of
near-solar metallicity.  The irregularities remaining after continuum
flattening occur over a sufficiently broad wavelength scale that the
indices defined here will not be affected appreciably. This feature
was of course a defining requirement in our development of this
method.

The data reduction process, given digital data, has the effect
that not all pixel boundaries are identical from spectrum to spectrum,
and nor will there necessarily be identity between pixel boundaries
and the boundaries of the theoretically-defined indices. We obviate
potential difficulties here by ensuring that the intensity
in each wavelength interval is calculated directly by integration of
the data, with the assumption that the intensity is constant over
wavelength inside any pixel which straddles the edges of the indices.
Errors introduced by this (common) assumption were found to be
negligible.

An absorption line index may be represented in terms of the ratio of
the sum of the fluxes in absorption bands to the sum of those in
nearby comparison bands, normalised so that the index has a value
of unity in the absence of absorption. To avoid problems arising from
large-scale errors in the intensity normalisation of the spectrum,
all wavelength bands of an index are required to lie within a
$\rm{few} \times 10\,${\AA} of each other.

The index is therefore defined to be
$$          I \;\; =  \;\;
    {  \; ( \; \sum_{i=1}^{N_C} \,\, {\Delta \lambda}_{C\,i} \; ) \;\;
      \sum_{i=1}^{N_A} \; F_{A\,i} \;  \over
      \; ( \; \sum_{i=1}^{N_A} \,\, {\Delta \lambda}_{A\,i} \; ) \;\;
    \sum_{i=1}^{N_C} \; F_{C\,i} \; }
$$
where $F_{X\,i}$ is the flux in the $i^{{th}}$ band, these being of
width $\Delta \lambda_{X,i}$, and $N_X$ is the total number of flux
bands, with $X=A$ indicating absorption bands, and $X=C$ indicating
comparison bands.

The values of each of the abundance and gravity indices of Tables 1
and~2 are then calculated for each object of interest, providing the
basic observational data from which the metallicity is estimated.
Similarly, a set of index values can be computed from the synthetic
fluxes.  These 16 synthetic indices may be calculated for each of the
100 different stellar models, potentially giving a grid of 1600 data
values.  Therefore each of the 100 points in the temperature --
gravity -- metallicity ($T_{ef\!f}$ -- $\log\,g$ -- [Fe/H]) parameter
space has a set of 16 index values associated with it.

\subsection {2.3 Iron Abundance -- Gravity Relations from Indices}

As detailed in \S 3 below, a temperature can be found independently of
these line strengths from photometry.  Given the effective
temperature, one calculates synthetic index values for each of the 16
indices of interest at each point in a grid in a metallicity --
surface gravity plane corresponding to the temperature of the
star. This may be accomplished by interpolation over temperature for
each set of points in the $T_{ef\!f}$ -- $\log\,g$ -- [Fe/H] parameter
space defined by a pair of gravity and metallicity values.
The three-dimensional $T_{ef\!f}$
-- $\log\,g$ -- [Fe/H] parameter space therefore collapses to a
two-dimensional [Fe/H] -- $\log\,g$ space, with 16 index values being
associated with each of the 20 points in the plane.

It now remains to determine numerically the metallicity--gravity
relations for each observed index.  Given an effective temperature,
the value that a particular index may have is constrained to lie in a
surface in the three-dimensional index -- metallicity -- gravity
parameter space. If the
value of the index $I$ appropriate to that star is determined from
observations, the line of intersection between the surface and the
plane defined by the observed index can be found. This represents the
[Fe/H] -- $\log\,g$ relation appropriate to the observed index and
temperature of the star.
By appropriate interpolations within this three-dimensional parameter
space, it is possible to calculate sufficient numbers of
([Fe/H],$\,\log\,g$) data points to define the [Fe/H] -- $\log\,g$
relation belonging to a particular observed index value, corresponding
to the temperature of the star.

These computational procedures could be repeated for each index,
establishing a different ([Fe/H]--$\log\,g$) relation from each.
Consequently, a set of 16 relations could be derived. These could be
used in principle to solve for the iron abundance and the gravity.

The indices adopted for the analysis were selected on the basis of
their sensitivities to metallicity, to gravity, or to both. They may
be classed into three broad categories, depending on their relative
sensitivities to iron abundance and/or to surface gravity.  In the
absence of  errors in the observational data or in the theoretical spectra,
the relations in the [Fe/H] --
$\log\,g$ plane obtained from them would intersect at a point, which
would define the [Fe/H] and $\log\,g$ values of the star. However,
noise in the spectra complicates the analysis. Observational errors in
the indices propagate through into the [Fe/H] -- $\log\,g$ relations,
causing them to meet not at a single point but rather in a region of
the [Fe/H] -- $\log\,g$ plane, with enclosed area dependent on the
noise in the spectrum.  The values of the abundance and gravity
corresponding to the centre of this region provide the best estimates of
the stellar parameters.  For the analysis process, the relations could
be plotted in a graph of $\log\,g$ against [Fe/H]. The interactive use
of a cursor was adopted as a means of finding the best intersection
point based on an eye estimate.  Although possessing some subjective
qualities, this procedure has the advantage that checks for gross
errors can be made, a helpful feature while developing the method,
while also allowing the rejection of apparently low-weight data
points.

However, the purpose of these analysis methods is the interpretation
of low-signal spectra. In this case, the noise in the individual index
values will be so large that it becomes difficult to define a solution
for iron abundance and for gravity. We therefore combine sets of individual
indices having similar sensitivities to abundance and gravity to form
{\it compound indicators.}

A compound indicator is defined in terms of single indices
$I_1 \, , \, I_2 \, , \, ... \, , \, I_{N_I} \, ,$ as
$$     C  \; \equiv \;  \sum_{j=1}^{N_I} \; \omega_j \, w_j \, I_j
$$
where $C$  is the compound indicator,
$N_I$ is the number of individual indices used, $I_j$ is the value of
the $j^{{th}}$ index used to define the compound indicator, $\omega_j$
is the weighting factor given to the $j^{{th}}$ index, in general
inversely proportional to the square of the error in that index, and
$w_j$ is a possible additional weighting factor (see Paper I for
details).

Three of these compound indicators are composed of indices sensitive
to metallicity only, one is made from indices sensitive to both
metallicity and gravity, and the final two are made from ionic gravity
indices. Provided that the
spectrum is not too badly affected by noise, the intersection
point of these six indicators in the [Fe/H]--$\log\,g$ plane
defines the iron abundance and the surface gravity of a star of interest.

We also defined a single comprehensive compound indicator which could
be used to determine the abundance of those stars whose spectra were
so noisy that the cursor method could not be applied. This required
that a surface gravity be assumed {\it a priori.}.  A single indicator
was therefore defined from all eleven indices of Table~1, by giving
the metallicity-only sensitive indices additional weighting factors
$w_j = 2\,,$ while the metallicity-gravity indices were given weights
of $w_j = 1\,.$ This weighting scheme was adopted to reduce the
gravity sensitivity of the indicator, while still incorporating those
indices which measured very strong absorption lines.

The definition of a second single indicator which combines only those
indices which are {\sl insensitive} to gravity can, when compared to
the metallicity- and gravity-sensitive compound indicator, provide
limited surface gravity information about the stars in the sample. A
second comprehensive compound indicator was therefore defined, using
the seven gravity-insensitive indices listed in Table~1, with no
extra weighting factors.

\goodbreak
\section{3. CALIBRATION OF }
\centerline{\bf THE  EFFECTIVE TEMPERATURE --  $(V-I)_{_C}$ RELATION}

Our analysis techniques require that the effective temperature be
specified as an independently determined parameter. This has the
advantage of reducing the number of parameters which are being
determined, allowing chemical abundance information to be obtained
from even lower signal spectra. We now establish a calibration of
effective temperature from $V-I$ photometry.

The analysis techniques were developed primarily to study stars at
high Galactic latitudes, along lines-of-sight for which the interstellar
reddening is small. In these cases a single color index measurement
is sufficient to provide an accurate effective temperature.  Although
the $(B-V)$ index shows a strong sensitivity to temperature, it also
shows a sensitivity to stellar metallicity and, to a lesser extent, to
luminosity, reducing its value as a temperature indicator.  For
example, by interpolation within the data of B\"{o}hm-Vitense (1981)
to an effective temperature of $T_{ef\!f} = 5000\,\rm{K}\,,$ a solar
metallicity dwarf has $(B-V)$ = $0{^m}{.}94\,,$ a solar metallicity
giant has (B$-$V) = $0{^m}{.}91\,,$ and a metal-poor dwarf,
$\log(Z/Z_{\odot}) =-1$, has $(B-V)$ = $0{^m}{.}78\,.$

In contrast, the Cousins' $(V-I)_{_C}$ (Cousins 1976; Bessell 1979)
index combines a temperature sensitivity with at most a slight dependence
on luminosity and metallicity.  As noted by Bessell (1979) for G and K
stars, for a given temperature there is little difference in
$(V-I)_{_C}$ between luminosity classes III and V. (The dwarf star
$(V-I)_{_C}$ -- $T_{ef\!f}$ relation determined below will be
assumed to apply to all luminosities.) An investigation by Johnson
{\it et al.} (1968) suggests that any changes with metal abundance in
the difference in line blanketing effects between the $V$ and $I_{_C}$
bands are likely to be small, producing only a very slight dependence
of the $(V-I)_{_C}$ index on metallicity. We confirm this below.

Bessell (1979) provides temperature calibrations for the
$(V-I)_{_C}$ and $(R-I)_{_C}$ indices for solar metallicity stars,
but does not consider metal-poor dwarfs.  We have therefore derived
the $(V-I)_{_C}$ -- $T_{ef\!f}$ -- [Fe/H] relation from
published observational data.

\subsection{3.1  Temperature Data}

Effective temperature measurements for late-type stars are available
in the literature, including information obtained by comparing
spectrophotometric data with model atmosphere predictions, from the
detailed analysis of stellar spectra, and using the infrared flux
method (Blackwell and Shallis, 1977; Blackwell, Shallis and Selby,
1979).  The infrared flux method, which involves comparison of
observed fluxes at infrared wavelengths with the integrated stellar
flux, has the advantage that it provides results which are insensitive
to uncertainties in model atmospheres, although the method is
dependent on the absolute calibration of stellar fluxes which is
adopted.  This method has been used by Saxner and Hammarb\"{a}ck
(1985) to study the temperature scale of F and early G dwarf stars.
Measurements were presented for 31 stars with metallicities close to
that of the Sun.  The stars are in general slightly hotter than is
required for a G dwarf calibration, with $T_{ef\!f} > 5800\,\rm{K}$
for all but one.  Magain (1987) extended the Saxner and Hammarb\"{a}ck
work to metal-poor stars, presenting temperatures for 11 stars in the
range $5300\,\rm{K} \leq T_{ef\!f} \leq 6150\,\rm{K}\,,$ $-2.3 \leq
\rm{[Fe/H]} \leq -1.2\,.$ These two sets of results provide accurate
data for the study of early G dwarfs, but they do not cover the entire
temperature range required for the F/G stars of interest.

A large body of temperature estimates  has been obtained by Peterson
and Carney (1979), by comparing observed spectrophotometric
distributions with a series of model-atmosphere predictions. They
obtained data for 74 dwarfs in the range $T_{ef\!f}$ = 4760 to 6300~K,
for a wide range of metallicities, in order to perform a calibration
of the Johnson $(R-I)_{_J}$ and $(V-K)$ indices. Carney (1983b)
provided revisions and additional results, including photometry for
the stars. Although the results are model dependent, they form a very
useful secondary source of data for the calibration of the
$(V-I)_{_C}$ index.

There are eight stars with temperature estimates both from the
infra-red flux method and from Peterson and Carney, allowing a
consistency check to be made.  The mean difference in temperature (in
the sense Peterson and Carney minus the infrared flux method) is 16~K,
with an root-mean-square difference of 51~K.  Thus the two methods
indeed yield consistent results and may be combined to construct a set
of temperature measurements for the calibration, retaining the
temperature estimate from the infrared flux method if
available.\footnote{$^3$}{One should bear in mind, however, that King
(1993) argues that this temperature scale is too low, by 150--200K,
for halo stars.}

\subsection{3.2  Photometric Data}

Photometric observations on the Cousins $VR_{_C}I_{_C}$ system were
made by Dean (1981) for some metal-poor stars. Of these, 21 are found
to have temperatures from the sources described above. For a few
stars, photometry by Taylor (1986) is available. Carney (1983b)
published extensive photometric results for metal-poor stars,
including most of those having Peterson and Carney temperatures, but
the VRI observations have been made on the Johnson
system. Fortunately, Carney (1983a) provided a transformation equation
relating his $(V-I)_{_J}$ index to Dean's $(V-I)_{_C}$ results,
valid over virtually the entire temperature range of interest. This
allows  Cousins $(V-I)_{_C}$ indices to be calculated from Carney's
extensive data, providing a secondary reference source of photometric
data.  According to Peterson and Carney (1979) and to Carney (1983a),
all stars for which they gave temperatures have negligible
reddening. Any additional stars are also likely to be sufficiently
close to the Sun that reddening can be ignored. The observed color
indices are therefore taken to be the intrinsic values for all the
stars of interest.

\subsection{3.3 Metallicity Data}

Metallicity estimates are given for many of the stars in the {\it
Catalogue of [Fe/H] Determinations} (Cayrel de Strobel {\it et al.,}
1985). Where available, a mean value of the [Fe/H] parameter was
calculated from the determinations listed in the {\it Catalogue,}
although the scatter is often quite large.  Fortunately the quality of
the [Fe/H] data is not critical, as the metallicity dependence of the
$(V-I)_{_C}$ index is expected to be small. If no spectroscopic
[Fe/H] estimate were available from the {\it Catalogue,} photometric
estimates were derived from the Peterson and Carney
$\delta({U}-{B})_{_{0.6}}$ color excesses using the calibration
of Carney (1979), or, when these were not given, color excesses were
calculated from the Carney (1983b) UBV photometry and normalised to
give the $\delta({U}-{B})_{_{0.6}}$ parameter by the method of
Sandage (1969).  Some additional photometric estimates have been taken
from the work of Norris {\it et al.} (1985).

The data collated from the literature are presented in Table~4,
listing star identification, $(V-I)_{_C}$, effective
temperature and metallicity.  Hyades stars were assigned a common
metallicity of [Fe/H] = $+0.12\,.$ A cross in the `rejection' column
indicates that that star was unsuitable for use in determining the
$(V-I)_{_C}$ -- temperature calibration.

The quality of the data from the various references varies
considerably.  Estimates of the stellar effective temperature based on
the infrared flux method were favored when available; for color
indices, actual $(V-I)_{_C}$ observations were used instead of
values calculated from the $(V-I)_{_J}$ index, if possible; for iron
abundances spectroscopic measurements were preferred. The orders of
preference for the data are summarised in Table~5.
A mean was calculated
if two or more values of the same reliability were available.

The effective temperature data are plotted against the $(V-I)_{_C}$
color index in Figure~1 in the form of $\theta_{ef\!f}$ parameter
($\equiv5040\,{\rm{K}}\,/\,T_{ef\!f}$), with different metallicity
ranges identified.  The data are distinguished by [Fe/H].  The trend
which is apparent in the Figure may be represented by a linear
relation of the form $$ \theta_{ef\!f} \; = \; a \; + \; b \;
({V}\!-\!{I})_{_C} $$ where $a$ and $b$ are constants.  An
unweighted least squares fit to the data is shown in the Figure, being
$$\eqalign{ \theta_{ef\!f} \; = \; &\;\;\;0.484 \;\;\;+ \; \;\;\;0.581
\, ({V}\!-\!{I})_{_C} \cr &\pm0.010 \qquad \pm0.014 } \eqno(1)
$$ Note that consistent values of the fit parameters are obtained by
fitting separately those data obtained from the infrared flux method
and those from spectrophotometric scans.  If the $\theta_{ef\!f}$ --
$(V-I)_{_C}$ data are divided into two subsets according to
metallicity, unweighted least squares fits to each give $$ \eqalign{
\theta_{ef\!f} = &\;\; 0.491 \! \qquad + \qquad \! 0.569 \,
({V}\!-\!{I})_{_C} \cr &\pm0.013 \qquad \pm0.019 } \eqno(2) $$
for [Fe/H]$ > -1.0$, and $$
\eqalign{ \theta_{ef\!f} = &\;\; 0.479 \!  \qquad + \qquad 0.593 \,
({V}\!-\!{I})_{_C} \cr &\pm0.016 \qquad \pm0.022 } \eqno(3) $$
for [Fe/H]$ \leq -1.0$.

These two relations are also consistent with each other to  the quoted errors,
implying no significant metallicity dependence.  Attempts to fit other
functional relations which include a [Fe/H]--dependence also fail to
reveal any clear sensitivity to [Fe/H].  We conclude that the
$(V-I)_{_C}$ -- temperature relation is insensitive to
stellar metallicity and is well approximated by Equation~(1).  This
relation is valid over the range $T_{ef\!f} = 4500\,\rm{K}$ to
$6700\,\rm{K}\,,$ $(V-I)_{_C}$ = 0.45 to $1.1\,,$ and has been
tested over [Fe/H] = $+0.1$ to $-2.7\,.$ The scatter of the data
points about the relation corresponds to $\pm 90\,\rm{K}$ at
$5500\,\rm{K}\,.$ An error of $\pm 0.01$ in $(V-I)_{_C}$ produces an
error in $T_{ef\!f}$ of $\pm 35\,\rm{K}$ at $5500\,\rm{K}$ (propagated
errors only).

A comparison of the temperature calibration of Equation~(1) with that
given by Bessell for solar-metallicity dwarfs reveals that his scale
is about $55\,\rm{K}$ hotter. The difference between his relation and
that here for [Fe/H] $ > -1.0$ (Equation~2) is typically only $45\,\rm{K}\,.$
Bessell based his calibration on the interferometric results of Code
{\it et al.} (1976) for early spectral types, which extend only to
late F-type stars, and on the occultation measurements for K and M
giants by Ridgway {\it et al.}  (1980). No data for G dwarf stars were
used. Although the calibration presented here does not use the
`direct' methods of temperature determination adopted by Bessell, it
has the significant advantage of exploring the parameter ranges of interest in
considerably greater detail.

Figure 2 shows a comparison of the relation of Equation (1)
with the work of Bessell, and with the theoretical calibration of the
$(V-I)_{_C}$ index given by VandenBerg and Bell (1985), which is based on
model-atmosphere and synthetic-spectrum computations.  For a given
$(V-I)_{_C}$ index, the model atmosphere results are hotter than
those of the present calibration by between $200\,\rm{K}$ to
$300\,\rm{K}$, for stars with [Fe/H] = $0.0\,,$ and by $150\,\rm{K}$ to
$200\,\rm{K}$ for [Fe/H] = $-1.0\,.$ The VandenBerg and Bell calculations
show a slight,
but clear, metallicity dependence in their photometric calibration,
amounting to $0{^m}{.}02$ in $(V-I)_{_C}$ between [Fe/H] = 0.0 and
$-2.0$, for stars having $T_{ef\!f} = 5500\,\rm{K}\,$ (see Gilmore, Wyse
and Jones, 1995). The difference
between the colors predicted by the metal-rich and metal-poor
calibrations of Equations (2) and (3) is $0{^m}.01$ for $T_{ef\!f} =
5500\,\rm{K}$ stars.
 There is therefore a small difference between
the temperature scale of VandenBerg and Bell and that of
Equation~(1).

\vfill\eject
\section{4. APPLICATION OF THE GENERAL METHOD}
\centerline{\bf  TO SPECIFIC SPECTRA}

\subsection{4.1 The Available Spectroscopic Data}

Spectra were obtained using the Autofib multi-fiber spectrograph of
the 3.9-metre Anglo-Australian Telescope, with the IPCS as detector, as
part of a radial velocity survey described by Wyse and Gilmore (1990).
A total of 60 working fibers were available for program objects and
blank sky (4 broken), over a field of view of $40\,$arcmin.  Sky
subtraction was achieved to about one percent accuracy, as described in
Wyse and Gilmore (1992).

The synthetic calibrations for the analysis techniques are available
for wavelength resolutions in the range 1.0 to 2.5{\AA} full-width at
half-maximum intensity.  The resolutions of the spectra were
determined from studies of the arc lamp calibration spectra. Unblended
lines were selected which were sufficiently strong that the profiles
could be measured accurately, but not so strong that the cores
suffered from detector saturation (when the photon arrival rate was so
high that the limited time resolution failed to distinguish between
successive photons).  The full widths at half-maximum intensity were
measured for fourteen lines across all wavelengths, independently for
spectra from optical fibers at the centre, and at the ends, of the
spectrograph slit. The observed wavelength range was
4025--4930{\AA}. For the first fiber on the spectrograph slit, the
mean full-width at half-maximum was $2.23\pm0.07\,${\AA} in the
4025--4150{\AA} region, $2.01\pm0.10\,${\AA} between 4500 and
4700{\AA}, and $2.13\pm0.20\,${\AA} between 4840 and 4950{\AA}.  The
mean value over all wavelengths was $2.11\pm0.07\,${\AA}.  For a
central fiber, the mean resolution over all wavelengths was
$2.12\pm0.04\,${\AA}, negligibly different.
The synthetic calibration nearest to this result,
that for $2.1\,${\AA}, was selected for all analyses of the
observational data.

\subsection{4.2 The  Analysis Technique for the 4025--4930{\AA} Region}

The general analysis described in Paper I defined optimum indices
between 4030 and 5000{\AA}. The available data demanded that the
analysis techniques be amended, to accommodate the more restricted
wavelength range available, and to use a smaller number of
spectroscopic indices. We emphasize here the implications of this
change, in order to illustrate the range of applicability of the present
technique.

There are nine abundance index flux bands in the list of Table~1 which
lie wholly or partly between 4930 and 5000{\AA}.  These were used
to define four abundance indices.  For present
application, a revised set of eight abundance indices was defined, as
presented in Table~6.
These eight indices formed
the basis of the analysis techniques as implemented in this Section,
for all the observed spectra.  Three of the original indices were
omitted in entirety (numbers 1, 2 and~4 of Table~1.  A fourth
index (number~5 of Table~1) was amended by deleting the
4925.8--4931.6~{\AA} comparison flux band.
Loss of the 4930--5000~{\AA} region
does not affect the ionic gravity indices and those listed in Table~2
were used for the work described below.

To solve for both iron abundance and surface gravity, using the restricted
4025--4930{\AA} region, a set of six compound indicators was defined from
the indices of Tables 6 and~2.
Three of these compound indicators are composed of indices sensitive
to metallicity only, one is made from indices sensitive to both
metallicity and gravity, and the final two are made from ionic gravity
indices. These are listed in Table~7. Provided that the
spectrum is not too badly affected by noise, it is the intersection
point of these six indicators in the [Fe/H]--$\log\,g$ plane which
defines the iron abundance and surface gravity of a star of interest.

Within the 4025--4930{\AA} region, the single comprehensive compound
indicator defined
from all indices sensitive to abundance used all eight indices of
Table~6. The single compound indicator defined from only those
indices which are {\sl insensitive} to gravity consisted of the five
gravity-insensitive indices listed in Table~6.

\subsection {4.3  Estimating signal-to-noise ratios }

As discussed in Paper~I, the accuracy of an abundance determination
depends strongly on the noise in the spectrum. In order to assess
the extent of the noise-induced [Fe/H] error,
signal-to-noise ratios were measured for the spectra from the
scatter of the individual pixel signals about their mean, applying
a correction to account for the expected contribution from line
absorption (Jones, 1991).

The signal-to-noise ratio per pixel $R_{S/N\,p}$ in a region of the
spectrum is given by
$R_{S/N\,p} \; = \; 1 \, / s_{noise} \,,$
where $s_{noise}$ is the ratio of the standard deviation of the
pixel fluxes contributed by noise in the spectrum to the mean flux.
However, due to the presence of line absorption, the observed
scatter in the pixel fluxes will be larger than that caused by
noise alone. On assuming that the effects of line absorption
and noise add in quadrature, the observed value $s_{obs}$ of the
ratio of the standard deviation to the mean pixel flux is given
by
$$
     s_{obs}^2  \; =  \; s_{abs}^2  \;  +  \; s_{noise}^2  \;\;, \eqno(4)
$$
where $s_{abs}$ is the ratio of the deviation contributed by line
absorption alone to the mean pixel flux. The parameter $s_{abs}$ was
determined from synthetic spectra, convolved with a representative
instrumental resolution profile. A table of $s_{abs}$ as a function
of stellar effective temperature and metallicity was generated.
Following an abundance analysis, the value of $s_{abs}$ appropriate
to any individual spectrum could be determined by interpolation
within these data and the value of $s_{noise}$ calculated
from the observed scatter of the pixel fluxes using Equation~(4).
Although tests using noise-added synthetic spectra confirmed the
accuracy of this approach, the method was used only if the noise
in the spectrum strongly dominated over the effects of line absorption:
no attempt was made to derive signal-to-noise ratios if
$(\, s_{obs} \, - \, s_{abs} \,) \,/\, s_{obs} > 0.15\,.$

The signal-to-noise ratios within the 4500--4690{\AA} region were
used to measure the quality of the spectroscopic data.
The ratio
of the standard deviation to the mean pixel flux was determined
in individual 50{\AA} sections and a mean value of $s_{obs}$
calculated for the entire 4500--4690{\AA} region,
to guard against large-scale irregularities
in the intensity normalisation contributing to $s_{obs}$.
 For a resolution
of 2.1{\AA} full-width at half-maximum, the values of  $s_{abs}$
range from 0.009 at $T_{ef\!f}=$6500K, [Fe/H]$=-2.0$ to
$s_{abs}= 0.32$ for solar-metallicity  dwarfs with temperature of 4500K.

\goodbreak
\section{5. APPLICATION AND CALIBRATION}
\centerline{\bf USING STANDARD STARS AND CLUSTERS}

Analysing observational data for stars of known chemical compositions
allows one to identify and isolate any errors in the calibration of
the abundance scale.  The synthetic spectra on which the present
technique is based were obtained by scaling a standard solar model
atmosphere, with the assumption of solar element ratios, which may
result in a systematic offset requiring a recalibration, particularly
for analyses of stars differing significantly from the Sun in terms
of metallicity, temperature and surface gravity.

Tests were performed using spectra of the twilight sky (essentially
the solar spectrum), of a selection of stars in the open cluster M67 and
of a sample of field stars known to cover a wide range in metallicity.
Photometrically-determined effective temperatures were obtained prior
to the analysis of the stellar spectroscopic data.

\subsection{5.1 Analyses of Twilight Sky Spectra}

Eight separate sets of 60 twilight sky spectra were analysed,
providing a zero-point for the abundance scale.  The individual
spectra have low signal-to-noise ratios, generally in the range 3 to
18 (for 1.0{\AA} wide bins). Following standard data reduction, the
480 separate twilight sky spectra were coadded to produce a low-noise
solar spectrum, which was analysed using both of the comprehensive
compound indicators introduced above.

A standard solar surface gravity of
$\log_{_{10}}(\,g\,/\,\rm{cms}^{-2}) = 4.44$ was adopted.  The
all-metallicity index comprehensive compound indicator, which includes
gravity-sensitive indices, yielded an iron abundance of [Fe/H] =
$-0.14$ for the co-added data, while the metallicity-only index
comprehensive compound indicator gave [Fe/H] = $-0.06\,.$ Thus there
is a small error in the zero-point of the abundance scale.

There are several conceivable causes of small errors in the
abundance scale. Given the accuracy of the techniques for the reduction
of synthetic spectra, numerical errors in the solution for the stellar
parameters are unlikely to be so large so as to account for these zero-point
problems. Neither should the inaccuracies introduced by the normalisation
of the continuum level be on this scale. Deficiencies will be present in the
synthetic spectra, due to the use for example of inaccurate $gf$-values.
However, given that the majority of oscillator strengths were
computed from the solar spectrum, it is to be expected that these
deficiencies will be minimised for solar analyses. The neglect of molecular
lines in the computation of the synthetic spectra would tend to
underestimate the absorption in both the absorption and comparison
bands of the indices, leading to a [Fe/H] result which would be too
large, in contrast to the underestimate actually obtained.
Similarly, the neglect of large numbers of weak metallic lines
not included in the Moore {\it et al.} compilation would cause an
overestimate. That the error of the all-metallicity indicator
is greater than that of the metallicity-only indicator might suggest
the presence of a problem which is worse for stronger Fe~I lines,
such as an inability to represent their damping wings correctly.
However, again the computation of oscillator strengths from solar
data would avoid such difficulties; only the lines synthesised using
published weighted oscillator strengths would be susceptible to
this.

The presence of scattered light, which is removed from program stars
by the sky subtraction procedure, is also unlikely to be an important
source of zero-point error, but will contribute to the random error.
As discussed in more detail by Wyse and Gilmore
(1992), for the setup typically used for the data aquisition, only
$\sim 4\%$ of the flux from one spectrum is scattered into nearby
spectra.  Uniform scattering adds only $0.2\%$ of the total flux in
an image to any one spectrum.

The coadded twilight sky/solar spectrum was then analysed using the
six separate compound indicators of Table~7. While the four iron
abundance sensitive indicators show only small scatter, the two ionic
equilibrium indicators show less consistency. Indeed, a difficulty
arises in the use of the ionic indicators in spectra having
resolutions as low as the 2.1{\AA} value of the available
observational data.  The ionic indices were optimised for use in the
analysis of spectra having resolutions close to 1.0{\AA}. The flux
bands frequently contain only single ionic absorption lines.
Consequently the indices are less sensitive to surface gravity for the
coarser resolutions considered here. Of the five ionic indices of
Table~2, tests with synthetic data reveal that index number 9 does
lose sensitivity to gravity as the resolution is degraded, becoming of
little use in analyses at a 2.1{\AA} resolution. The other four
indices do, however, continue to provide useful information at this
resolution.

Having analysed the coadded data, the 480 individual solar spectra
were reduced and iron abundances were determined from each.  On
assuming the solar surface gravity, the mean iron abundance result
from the all-metallicity comprehensive compound indicator was [Fe/H] =
$-0.17$, with a standard deviation of 0.23, while that from the
metallicity-only comprehensive compound indicator was [Fe/H] =
$-0.10$, with a standard deviation of 0.25 dex.
These are in satisfactory  agreement with the results from the
higher signal co-added data.

The analyses of the twilight sky data have shown consistently that
there is a small zero-point error in the abundance results. This is
corrected for by adjusting the output of the all-metallicity
comprehensive compound indicator by adding $+0.14$ dex, while those of
the metallicity-only comprehensive indicator were adjusted using a
correction of $+0.06$ dex.  This practice was adopted for all later
abundance results.

\subsection{5.2 Analysis of Spectra of Stars in the Cluster M67}

A sample of 74 stars in the old open cluster M67 was used for further
tests of the abundance analysis techniques.  This cluster offers a
selection of well-studied stars which have  published photometry by several
different authors. Being of the same chemical composition, these stars
enable the accuracy of the techniques to be investigated as a
function of temperature, and, to some degree, of surface
gravity. Spectra were obtained in May 1989 for a number of dwarfs,
turn-off stars and subgiants.  Cluster stars are known by a variety of
different designations, of which those of Eggen and Sandage (1964),
Racine (1971) and Sanders (1977) are most common.
The Racine naming system is an extension of that of Eggen and Sandage,
and is adopted here. Signal-to-noise ratios in 1.0{\AA} wide
bins were in the range  5 to 27, with a
majority between 9 and 19.

Cross-correlations of the M67 spectra with standard templates provided
the radial velocities which are required for the correction of the
wavelength scales to the laboratory rest frame. These radial velocity
data also provided a means of testing cluster membership, in addition
to identifying some binary star systems.  Non-binary members of the
cluster should all have similar radial velocities, since the velocity
accuracy for our spectral resolution and wavelength range, 5--10 km/s,
is considerably larger (by an order of magnitude) than the intrinsic
velocity dispersion of the cluster.

An initial test of the suitability of a star for abundance analysis
was therefore made by rejecting those having velocities outside a
narrow range centred on the mean radial velocity of the M67 sample.
The distribution of velocities of the cluster members was well fitted
by a Gaussian function having a $1/\sqrt{\rm{e}}$-half-width of
$\sigma = 6\,\rm{kms}^{-1}\,,$ which is dominated by the random errors
in the cross-correlation results, together with non-Gaussian wings due
to non-members and spectroscopic binaries.  A membership criterion
that a star had to have a radial velocity within $2\sigma$ of the
cluster mean produced a sample of 62 cluster members and 12
non-members/binaries. Given the relative frequencies of the background
and Gaussian velocity distributions, there is a $95\,\%$ probability
that any given star classified as a member by this criterion is indeed
a genuine member of the cluster.

Precision radial velocity results are available in the literature for
some M67 stars, allowing a check to be made on the reliability of our
adopted membership criterion.  Mathieu {\it et al.} (1986) give such
data for 33 stars from the present sample of 74.  Of these 33 stars,
26 passed a $2\sigma$ velocity test using a mean velocity for the
cluster stars and the Mathieu {\it et al.} error estimates. One of
these was rejected as being a spectroscopic binary using the results
of Mathieu {\it et al.} (1986, 1990).

Sanders (1977) performed a proper motion study of stars in the region
around M67, providing membership probabilities.  Similarly, Girard
{\it et al.} (1989) presented membership probabilities based on proper
motions and distances from the cluster centre.  The probability
distributions of both surveys are strongly bimodal, with clear peaks
at low and high probabilities. A 50\% membership probability test was
applied to the M67 stars having spectroscopic data. It should be noted
that because of the highly bimodal distribution, this 50\% criterion
conservatively rejects non-members, without leaving contamination
problems. For a star to be regarded as a cluster member, it had to
pass the membership tests using probabilities from both proper motion
surveys and also had to pass the radial velocity tests.  Our final
sample of M67 members contained 44 stars.

Photometry for M67 stars is available from a number of sources,
allowing effective temperatures to be calculated with reasonable
accuracy.  $BV$ data were taken from the studies of Eggen and Sandage
(1964), Murray {\it et al.} (1965), Racine (1971), Schild (1983),
Sanders (1989) and Gilliland {\it et al.} (1991).  The data were
obtained using a variety of techniques. For example, Gilliland {\it et
al.} and Schild used CCD photometry, the results of Eggen and Sandage
and the new data of Sanders were photoelectric, while Murray {\it et
al.} and Racine used photography.  It might be expected that the
photographic results would be of lower quality. To test this, the
$({B}-{V})$ data of Racine were compared with those of Gilliland
{\it et al.} for a sample of 23 stars in common to both studies. The
mean difference (in the sense Racine minus Gilliland {\it et al.}) was
$-0^m.012\,,$ with a root-mean-square difference of $0^m.047\,.$
Therefore, if available, the photoelectric and CCD data were
preferred, with mean $(B-V)$ results being calculated if
data were available from more than one photoelectric or CCD source. If
only photographic photometry was available, the color index was taken
from Racine, or if unavailable, from Murray {\it et al.}

Published photometry exists for M67 stars for other color indices.
Cousins $(V-R)_{_C}$ and $(R-{I})_{_C}$ data were
taken from Gilliland {\it et al.} (1991), Taylor and Joner (1985, 1988)
and Janes and Smith (1984). $(b-{y})$ and $\beta$ data were
taken from Nissen, Twarog and Crawford (1987), Anthony-Twarog (1987)
and Taylor (1978).

The $(B-V),$ $(V-R)_{_C}$, $(R-{I})_{_C}$
and $(b-{y})$ data were corrected for interstellar reddening
by assuming the $(b-{y})$ color excess of $E_{(b-y)} =
0^m.023$ measured using precision uvby$\beta$ photometry by Nissen,
Twarog and Crawford (1987). This $(b-{y})$ color excess was
converted to the $(B-V)$ excess using the
$E_{(b-y)}/E_{(B-V)}$ ratio of Crawford (1975).
Similarly, the $(V-R)_{_C}$ and $(R-\rm{I})_{_C}$
excesses were calculated from the $(B-V)$ excess using
data from the Savage and Mathis (1979) interstellar extinction curve.
The color excesses adopted were $E_{_{(B-V)}} = 0^m.032\,,$
$E_{_{(V-R)}} = 0^m.015$ and $E_{_{(R-I)}} = 0^m.022\,.$

A number of different color indices were therefore available for the
calculation of effective temperatures. Unfortunately, the available
photometric data were sufficiently diverse and disparate that it was
not feasible to achieve empirical temperature calibrations from the
literature. Insufficient published data were available for M67 stars
to define transformations between the different colors and the
$(V-I)_{_C}$ index for the calibration of Equation~1 to be used.
Instead, theoretical temperature calibrations were used for each of
the available color indices. Surface gravities were computed for the
stars of interest from published $V$-band apparent magnitudes.  The
$V$ apparent magnitudes were converted to absolute magnitudes using a
distance modulus of $9^m.71$ (Nissen, Twarog and Crawford
1987). Interpolation within the theoretical isochrones of VandenBerg
(1985) for dwarfs, and of Green, Demarque and King (1987) for
subgiants, provided the surface gravities corresponding to the
absolute magnitudes.  Cubic spline interpolation within the synthetic
color index calibrations of Bell and Gustafsson (1979, 1989) and of
VandenBerg and Bell (1985) gave the effective temperatures appropriate
to the dereddened photometry, the calculated gravities and an assumed
[Fe/H] of 0.0. An unweighted mean effective temperature was calculated
from the individual estimates of each color index.
The temperature
scale adopted out of necessity for the M67 stars is therefore that of
the Bell theoretical calibration, in contrast to the empirical scale of
Equation~(1). Figure~2 indicates a difference of about 200K between
the scale of Equation~(1) and the model atmosphere calibration of
the $(V-I)_{_C}$ index; adopting a temperature scale 200K cooler
would lead to iron abundance measurements for M67 stars 0.16dex
smaller.

The spectra were prepared in the standard manner; after the
subtraction of the night sky spectrum, the continua were normalised to
a constant level and the velocity corrections were applied. The
spectra were analysed using both the all-metallicity and the
metallicity-only comprehensive compound indicators, as amended for the
4025--4930{\AA} region. The 2.1{\AA} resolution synthetic flux
calibration was adopted.  The photometric temperatures and the
gravities derived from absolute magnitudes were used in the analyses.
The zero-point corrections to the [Fe/H] results derived from the
twilight sky spectrum were applied.

The iron abundance results provided an excellent test of the
consistency of the analysis methods during a study of a sample of
stars having a common metallicity. It was possible to search for the
presence of systematic errors as a function of temperature.  However,
the interpretation of the results is complicated by the relatively low
signal-to-noise ratios of the spectroscopic data, even by the
standards of the methods developed here.  Adopting [Fe/H] values from
those spectra having signal-to-noise ratios (in 1.0{\AA} bins)
$R_{_{S/N\,1{\AA}}} > 11\,,$ the mean result was $\rm{[Fe/H]} = -0.15$
from the all-metallicity comprehensive compound indicator, with an
internal error in the mean of $0.03$, with $\sigma_{[Fe/H]} = 0.16$.
Of the 45 successful [Fe/H] results, 43 had signal-to-noise ratios
above this limit.  The metallicity-only indicator gave a figure of
$\rm{[Fe/H]} = -0.14 \pm 0.03$, $\sigma_{[Fe/H]} = 0.18$, for the same
43 spectra.  We therefore adopt $\rm{[Fe/H]} = -0.14 \pm 0.10$ as the
iron abundance of M67, where the error estimate attempts to account
for all error sources, random and systematic.

Figure~3 shows the [Fe/H] results from the all-metallicity indicator
plotted against temperature for the subgiants and main-sequence
turn-off stars. There is no clear evidence of any temperature
dependence in the [Fe/H] results.  Figure~4 gives an equivalent plot
for the dwarf stars.  The dwarfs, however, do not show as constant a
result with effective temperature as do the subgiants; the coolest
dwarf stars, those with $T_{ef\!f} \simlt 5300$K do tend to have
lower [Fe/H] estimates.  A very similar result was obtained by Carney
{\it et al.} (1987; see their Figure 4) in their analysis of low
signal-to-noise, high resolution spectra by comparison with synthetic
spectra based on Kurucz models.  Those models, similarly to the
present models, did not take proper account of molecular line
absorption, which may be expected to be most important in the cooler
stars.  Thus any
 [Fe/H] dependence on
temperature seen  Figure~4 may well likely be a model artefact.

M67 has had a controversial history of chemical abundance and
reddening measurements (Taylor, 1982), with older studies finding
metallicities in the range from metal-poor to super-metal-rich.
However, in recent years a consensus has been achieved which finds the
cluster to have a near-solar metallicity. A selection of recent
abundance data is presented in Table~8; the final [Fe/H] result for
M67 from this study, $\rm{[Fe/H]} = -0.14 \pm 0.10$, compares
favorably with these results.

This implies that the iron abundance analysis techniques presented
here are able to provide accurate results from low-signal spectra.

\subsection{5.3 Analyses of Spectra of Field Standard Stars}

Spectra of a sample of 25 field stars having published metallicity data
were obtained in May 1987 and October 1988 using the Autofib system, in a
similar manner to the M67 stars.  These were selected from the list
used by Friel (1987) and from the set of metallicity estimates by
Laird, Carney and Latham (1988; hereafter LCL).  The six stars considered by
Friel had
published high-resolution analyses. Of these, three were dwarfs and
three were giants.  The other 19 stars, having between them 21
individual spectra, had metallicities from LCL,
and are expected to be dwarfs on the basis of their selection
criteria.  These 25 objects cover the [Fe/H] range from $+0.4$ to
$-2.9\,.$ Signal-to-noise ratios (for 1.0{\AA} pixels) were in the
range $R_{S/N\,1{\AA}} = 20$ to 90, higher than for the twilight sky
and M67 data.  The spectra were reduced in a similar manner to the M67
data.

To ensure consistency with the literature results, the same effective
temperatures used in the published analyses were employed.  The
spectra were analysed using the six compound indicators of Table~7. An
eye estimate provided the position of the best intersection point in
the [Fe/H] -- $\log g$ plane.  Each spectrum was then analysed using
the all-metallicity comprehensive compound indicator, and again with
the gravity-insensitive indicator. Iron abundances were determined by
specifying published surface gravities for the stars having
high-resolution analyses, or by assuming dwarf gravities for the LCL
stars.  The results are presented in Table~9 (stars with published
high-resolution abundance analyses) and in Table~10 (stars from LCL).
One analysis failed: that for G41-41 ([Fe/H] = $-2.9$) did not return
a [Fe/H] -- gravity relation, but this is understandable since the
star is well outside the metallicity range of our synthetic spectra,
being an order of magnitude more metal-poor than the extent of our
calculated grid.  In adition, the scatter of the [Fe/H]--$\log g$
relations for HD~184266 were too great for a best intersection region
to be defined using the six indicator cursor method.

The results obtained with the all-metallicity indicator are compared
with the published [Fe/H] data in Figure~5. The error bars in the
figure represent the errors estimated from the signal-to-noise ratios
of the spectra (see section 4.3 above).  The result for HD~218857 is
exceptional in differing substantially, by around an order of
magnitude, from the high-resolution study abundance by Luck and Bond
(1981). No explanation for this difference can be found.  G18-55,
which was observed twice, provided two abundance results which are
inconsistent with the LCL value.  The all-metallicity indicator
results for this star are not formally consistent : [Fe/H] = -0.46 +/-
0.08 and -0.61 +/- 0.04 . However, the metallicity-only, and six
indicator cursor method, results for the two spectra are
consistent. G18-55 is, however, a spectroscopic binary (see Carney
{\it et al.} 1994) and on this basis we chose not to use this star to
assess the accuracy of the analysis methods.

A comparison of the all-metallicity results from the remaining 23
spectra with the published data gave a mean difference in [Fe/H] of
$+0.06$ (in the sense this work minus literature value), with a
root-mean-square difference of 0.24.  For the 13 spectra having [Fe/H]
in the range 0.0 to $-1.2$ (the metallicity range of greatest interest
to these analysis techniques), the mean difference in [Fe/H] was
$-0.04$ and the root-mean-square difference was 0.13.  Excluding the
remaining known binaries from this comparison, {\it viz.} HD 149414
(Mayor and Turon 1982) and G18-28 (Carney {\it et al} 1994), changes
these numbers only slightly to a mean difference of $+0.07$ dex with
rms of 0.25 dex for all 21 stars, and for the stars with [Fe/H] in the
range $0.0$ dex to $-1.2$ dex the mean is unchanged, and the rms
increases to 0.14

There is some evidence that these techniques
overestimate abundances for very metal-poor stars ($\rm{[Fe/H]} <
-1.5$).  However, as the available data are limited for these very low
metallicities, we do not make any attempt to recalibrate the abundance
scale of the present techniques.  In general the results from this
work are in excellent agreement with the data from the literature,
which have been obtained from high-resolution spectroscopy.

\section {6. CONCLUSIONS}

The analysis presented above shows that spectra obtained from
fiber-fed spectrographs, despite their unpromising initial impression,
are indeed capable of providing a rather accurate estimate of the true
iron abundance, with uncertainly $\sim 0.2$ dex, given an input
estimate of the other important stellar parameters, namely the
effective temperature and surface gravity.  Thus the chemical history
of the Galaxy is amenable to study through large surveys of distant
stars, without unreasonable requirements  of telescope time.

\centerline  {ACKNOWLEDGEMENTS}
\bigskip
JBJ thanks Mike Edmunds for his advice and helpful discussions, and
acknowledges the support of an SERC studentship while this work was initiated.
The Center for Particle Astrophysics is supported by the
NSF.  RFGW acknowledges support from the AAS Small Research Grants
Program in the form of a grant from NASA administered by the AAS, from
the NSF (AST-8807799 and AST-9016226) and from the Seaver Foundation.
Our collaboration was aided by grants from NATO Scientific Affairs
Division and from the NSF (INT-9113306).  GG thanks Mount Wilson and
Las Campanas Observatories for access to their excellent facilities,
as a Visiting Associate, during the early stages of this work.

\vfill\eject

\parindent=0pt
\centerline {REFERENCES}
\medskip

\pp Anthony-Twarog, B. J., 1987. {  AJ}, { 93},
     647

\pp Bell, R. A., and Gustafsson, B., 1979. {  A\&AS},
 { 34}, 229

\pp Bell, R. A., and Gustafsson, B., 1989. { MNRAS},
       { 236}, 653

\pp Bessell, M. S., 1979. { PASP},
{ 91}, 589

\pp Blackwell, D. E., Petford, A. D., Arribas, S., Haddock, D. J.,
     and Selby, M. J., 1990. { A\&A}, { 232},
     396.

\pp Blackwell, D. E., and Shallis, M. J., 1977. { MNRAS}
     { 180}, 177

\pp Blackwell, D. E., Shallis, M. J., and Selby, M. J., 1979.
     { MNRAS}, 188, 847

\pp B\"{o}hm-Vitense, E., 1981. { ARA\&A}, { 19}, 295

\pp Burstein, D., Faber, S. M., and Gonzalez, J. J., 1986.
    {AJ},{ 91}, 1130

\pp Carney, B. W., 1979. { ApJ},{ 233}, 211

\pp Carney, B. W., 1983a. { AJ},{ 88}, 610

\pp Carney, B. W., 1983b. { AJ},{ 88}, 623

\pp Carney, B.W., Laird, J.B., Latham, D.L. and Kurucz, R., 1987. AJ, 94, 1066

\pp Carney, B.W., Latham, D.L., Laird, J.B. and Aguilar, L., 1994. AJ, 107,
2240

\pp Cayrel de Strobel, G., Bentolila, C., Hauck, B., and Duquennoy, A.,
     1985. { A\&AS},  { 59},
     145

\pp Code, A. D., Davis, J., Bless, R. C., and Hanbury Brown, R.,
     1976. { ApJ,} { 203}, 417

\pp Cousins, A. W. J., 1976. { Mem  RAS}, { 81}, 25

\pp Crawford, D. L., 1975. { AJ},{ 80}, 955

\pp Dean, J. F., 1981. { MNASSA }
 { 40}, 14

\pp Eggen, O. J., and Sandage, A., 1964. { ApJ},{ 140},
      130

\pp Fran\c{c}ois, P., 1986. { A\&A}, { 160},
      264

\pp Friel, E. D., 1987. { AJ},{ 93}, 1388

\pp Garcia Lopez, R. J., Rebolo, R., and Beckman, J. E., 1988.
     { PASP}, { 100}, 1489

\pp Gilliland, R. L., Brown, T. M., Duncan, D. K., Suntzeff, N. B.,
     Lockwood, G. W., Thompson, D. T., Schild, R. E., Jeffrey, W. A., and
     Penprase, B. E., 1991. { AJ},{ 101}, 541

\pp Gilmore, G., Wyse, R. F. G., and Jones, J. B., 1995.
      { AJ}, in press

\pp Girard, T. M., Grundy, W. M., L\'{o}pez C. E., and van Altena, W. F.,
    1989. { AJ},{ 98}, 227

\pp Gratton, R. G., 1989. { A\&A}, { 208}, 171

\pp Green, E. M., Demarque, P., and King, C. R., 1987. { The
     Revised Yale Isochrones,} Yale Univ. Obs., New Haven.

\pp Hearnshaw, J. B., 1975. { A\&A}, { 38},
     271

\pp Hobbs, L. M., and Thorburn, A., 1991. { AJ},
    { 102}, 1130

\pp Holweger, H., and M\"{u}ller, E. A., 1974. { Solar Physics},
      { 39}, 19

\pp Hoffleit, D., 1964. { Catalogue of Bright Stars,} Third
      Revised Edition, Yale University Obs., New Haven, Connecticut.

\pp Janes, K. A., and Smith, G. H., 1984. { AJ},{ 89},
     487

\pp Johnson, H. L., MacArthur, J. W., and Mitchell, R. I., 1968.
      { ApJ},{ 152}, 465

\pp Jones, J. B., 1991. PhD thesis, Univ. Wales, Cardiff.

\pp Jones, J. B., Gilmore, G., and Wyse, R. F. G., 1995, MNRAS submitted,
      (Paper~I)

\pp King, J.R., 1993. AJ, 106, 1206

\pp Laird, J. B., Carney, B. W., and Latham, D. W., 1988.
      { AJ},{ 95}, 1843 (LCL)

\pp Luck, R. E., and Bond, H. E., 1981. { ApJ},
     { 244}, 919

\pp Magain, P., 1987. { A\&A}, { 181},
     323

\pp Mathieu, R. D., Latham, D. W., and Griffin, R. F., 1990.
     { AJ},{ 100}, 1859

\pp Mathieu, R. D., Latham, D. W., Griffin, R. F., and Gunn, J. E., 1986.
     { AJ},{ 92}, 1100

\pp Mayor, M. and Turon, C. 1982. A\&A, 110, 241

\pp Moore, C. E., Minnaert, M. G. J., and Houtgast, J., 1966.
     { The Solar Spectrum $2935\,${\AA}, to $8770\,${\AA}, Second
     Revision of Rowland's Preliminary Table of Solar Spectrum
     Wavelengths,} Monograph 61, National Bureau of Standards,
     Washington D. C.

\pp Murray, C. A., Corben, P. M., and Allchorn, M. R., 1965.
      { Royal Obs Bull}, No. 91.

\pp Nissen, P. E., 1981. { A\&A},
     { 97}, 145

\pp Nissen, P. E., Twarog, B. A., and Crawford, D. L., 1987.
    { AJ},{ 93}, 634

\pp Norris, J. P., Bell, R. A., Butler, D., and Deming, D., 1980.
    { BAAS}, { 12}, 458

\pp Norris, J. E., Bessell, M. S., and Pickles, A. J., 1985.
     { ApJS}, { 58}, 463

\pp Peterson, R. C., 1980. { ApJ},{ 235}, 491

\pp Peterson, R. C., and Carney, B. W., 1979. { ApJ},
 { 231}, 762

\pp R. Racine, 1971. { ApJ},{ 168},
      393

\pp Ridgway, S. T., Joyce, R. R., White, N. M., and Wing, R. F.,
     1980. { ApJ},{ 235}, 126

\pp Sandage, A., 1969. { ApJ},{ 158},
      1115

\pp Sanders, W. L., 1977. { A\&AS}, { 27},
      89

\pp Sanders, W. L., 1989. { Rev Mex A\&A},
     { 17}, 31

\pp Savage, B. D., and Mathis, J. S., 1979. { ARA\&A},
 { 17}, 73

\pp Saxner, M., and Hammarb\"{a}ck, G., 1985. { A\&A},
 { 151}, 372

\pp Schild, R. E., 1983.  { PASP},{ 95},
      1021

\pp Taylor, B. J., 1978. {  ApJS}, { 39},
     173

\pp Taylor, B. J., 1982.  { Vistas in Ast},  { 26},
      253

\pp Taylor, B. J.,  1986. { ApJS},
     { 60}, 577

\pp Taylor, B. J., and Joner, M. D., 1985. { AJ},{ 90},
     479

\pp Taylor, B. J., and Joner, M. D., 1988. { AJ},{ 96},
     211

\pp van Altena, W. F., 1969. { AJ}, { 74}, 2

\pp van Bueren, H. G., 1952. { BA Inst Netherlands}, { 11}, 385

\pp VandenBerg, D. A., 1985. { ApJS}, { 58}, 711

\pp VandenBerg, D. A., and Bell, R. A., 1985. { ApJS}, { 58}, 561

\pp Wyse, R. F. G., and Gilmore, G., 1990, in { Chemical and
     Dynamical Evolution of Galaxies,} eds. F. Ferrini, J. Franco and F.
     Matteucci, ETS Editrice, Pisa, p19

\pp Wyse R.F.G. and Gilmore, G., 1992, MNRAS, 257, 1

\vfill\eject

\section {FIGURE CAPTIONS}

Fig. 1 :
The $\theta_{ef\,f}$ ($\equiv 5040{\rm K}/T_{ef\,f}$) temperature
parameter plotted against Cousins $(V-I)_{_C}$ color index for
the sample of stars of Table~4. Stars are distinguished by
metallicity~: vertical crosses denote stars having
${[Fe/H]} > -0.5,$ diagonal crosses stars having
$-0.5 \leq {[Fe/H]} > -1.5,$ and circles stars having
${[Fe/H]} \leq -1.5.$
The best fitting straight line through the data points
(for all metallicities) is shown.

Fig. 2 :
A comparison of the temperature scale of Equation~1 with the calibrations
of Bessell (1979) and of VandenBerg and Bell (1985). The relation of
Equation~1 (solid line) applies to dwarf stars of all metallicites, that
of Bessell (triangles) to dwarfs of solar metallicity, while those of
VandenBerg and Bell to dwarfs having [Fe/H] = $0.0$ (solid circles)
and $-1.0$ (open circles).

Fig. 3 :
The iron abundance results plotted against stellar temperature for
the objects of the M67 sample classified as subgiants and turn-off
stars. Stars are distinguished according to whether they are likely
cluster members or binaries (closed symbols for cluster members not
known to be binaries, open symbols for binaries and non-members).
They are also distinguished according to the source of the photometry
(squares for CCD or photoelectric photometry, triangles for photographic
photometry only).

Fig. 4 :
The iron abundance results plotted against stellar temperature for
the stars of the M67 sample classified as dwarfs. Stars are distinguished
according to whether they are likely cluster members and by the nature of
the photometry used to calculate effective temperatures, with symbols
defined as in Fig.~3.

Fig. 5 :
A comparison of the field star iron abundance results of this work with
published [Fe/H] data. The iron abundance results were obtained using the
single comprehensive indicator defined from all abundance sensitive
indices. Triangles denote stars having high-resolution analyses, while
squares represent those having only Laird, Carney and Latham (1988)
metallicities. Open symbols denote visual doubles or stars
which  are or are suspected of being spectroscopic binaries.
Solid symbols denote stars which are not known or suspected of being
double or binaries. Error bars represent the predicted noise-induced
error in the [Fe/H] results.

\bye